\documentclass[12pt,article,nofootinbib]{revtex4}
\usepackage{pdfpages}
\usepackage{epsf}
\usepackage{subfigure}
\usepackage{amsmath}
\usepackage{eqnarray,amsmath}
\usepackage{epstopdf}
\usepackage{mathrsfs}
\usepackage{natbib}
\usepackage{amssymb,latexsym}
\usepackage{dcolumn}
\usepackage{graphicx}
\usepackage{color}

\newtheorem{thm}{Theorem}[section]
\newtheorem{prop}[thm]{Proposition}%[section]
\def\be{\begin{equation}}
\def\ee{\end{equation}}
\def\ba{\begin{eqnarray}}
\def\ea{\end{eqnarray}}
\begin{document}
\title{\large \bf The spherical symmetry Black hole  collapse in expanding universe}
\author{J. T. Firouzjaee}
\affiliation{Institute for Gravitation and the Cosmos Physics Department, Penn State, University Park, PA 16802,
 USA and \\ Department of Physics, Sharif University of Technology,
 Tehran, Iran}
 \email{firouzjaee@physics.sharif.edu}

\begin{abstract}

The spherical symmetry Black holes are considered in expanding
background. The singularity line and the marginally trapped tube
surface behavior are discussed. In particular, we address the
conditions of whether a dynamical horizon forms for these cosmological
black holes. We also discuss the cosmological constant effect
on these black holes and the redshift of the light which comes from
the marginally trapped tube surface.

\end{abstract}
\pacs{95.30.Sf,98.80.-k, 98.62.Js, 98.65.-r}
\maketitle
\section{introduction}

The study of black holes in stationary and asymptotically flat spacetimes has led to many remarkable insights. But, as we know, our universe is not stationary and is in fact undergoing cosmological expansion. Let us use the term \emph{cosmological black hole} for any solution of Einstein equations representing a collapsing overdense region in a cosmological background, leading to an infinite density at its center \cite{sultana}. The first attempt in this direction is due to McVittie \cite{McVittie} who introduced a spacetime metric that represents a point mass embedded in a Friedmann-Robertson-Walker (FRW) universe. There have been several other attempts to construct solutions of Einstein equations representing such a collapsing central mass. Gluing of a Schwarzschild manifold to an expanding FRW manifold is one of such attempts, made first by Einstein and Straus \cite{Einstein Straus}. \\

Now, a widely used metric to describe the gravitational collapse of a spherically symmetric dust cloud is the so-called Tolman-Bondi-Lemaitre (LTB) metric \cite{LTB}. These models have been extensively studied for the validity of the cosmic censorship conjecture \cite{cencorship, joshi}. It was pointed out in \cite{man} that the model admits cosmological black holes. \\ 
	
Much of the literature on BH's focuses on stationary and asymptotically flat situations \cite{waldbook}. It is therefore desirable to have black hole models embedded in a cosmological environment to see if there may be considerable differences from the familiar Schwarzschild black hole and if there are quasi-local characteristics of it in place of the global and teleological event horizon. The need for a local definition of black holes and their horizons have led to concepts such as Hayward's trapping horizons \cite{Hayward94}, Ashtekar's isolated horizons \cite{ashtekar99}, Ashtekar and Krishnan's dynamical horizon (DH) \cite{ashtekar02}, and Booth and Fairhurst's slowly evolving horizon  \cite{booth04}. \\
										
 There have been some previous studies of BH's in cosmological situations, for example, to glue two different LTB metrics to study the structure formation out of an initial mass condensation or the formation of a galaxy with a central black hole \cite{kra-hel-sf} and \cite{kra-hel-BH}. However, since the structure of the metric outside these mass condensations are fixed by hand to match to a specific galaxy or cluster feature, we are faced with the shortcomings of the cut and paste models. Harada et. al was also interested in the behavior of primordial black holes within cosmological models to probe the gravitational memory and back hole radiation in expanding universe \cite{harada}.\\ Our main interest is to consider the black hole (BH) properties within the expanding universe model. We consider different properties of the LTB metric as a cosmological BH in section II. These properties are generalized to perfect fluid in section III. Section IV is devoted to the cosmological constant effect on spherical cosmological BH.

\section{LTB metric }
 The LTB \cite{LTB} metric is a
spherically symmetric non-static solution of the Einstein equations
with a dust source.

 The LTB metric may be written in synchronous coordinates as
\begin{equation}
 ds^{2}=-dt^{2}+\frac{R'^{2}}{1+f(r)}dr^{2}+R(t,r)^{2}d\Omega^{2}.
\end{equation}
and represents a pressure-less perfect fluid satisfying
\begin{equation}
\rho(r,t)=\frac{2M'(r)}{ R^{2}
R'},\hspace{.8cm}\dot{R}^{2}=f+\frac{2M}{R}.
\end{equation}
Here dot and prime denote partial derivatives with respect to the
parameters $t$ and $r$ respectively. The angular distance $R$,
depending on the value of $f$, is given by
\begin{eqnarray}\label{ltbe1}
R=-\frac{M}{f}(1-\cos(\eta(r,t))),\nonumber\\
\hspace{.8cm}\eta-\sin(\eta)=\frac{(-f)^{3/2}}{M}(t-t_{n}(r)),
\end{eqnarray}
for $f < 0$, and
\begin{equation}\label{ltbe2}
R=(\frac{9}{2}M)^{\frac{1}{3}}(t-t_{n})^{\frac{2}{3}},
\end{equation}
 for $f = 0$, and
\begin{eqnarray} \label{ltbe3}
R=\frac{M}{f}(\cosh(\eta(r,t))-1),\nonumber\\
\hspace{.8cm}\sinh(\eta)-\eta=\frac{f^{3/2}}{M}(t-t_{n}(r)),
\end{eqnarray}
for $f > 0$.\\
The metric is covariant under the rescaling
$r\rightarrow\tilde{r}(r)$. Therefore, one can fix one of the three
free functions of the metric, i.e. $t_{n}(r)$, $f(r)$, and $M(r)$.
One can shows that $M(r)$ represents the mass accumulation in a
2-sphere with
 radius $r$, more precisely, the Misner-Sharp mass \cite{manmass}.\\
There are two generic singularities of this metric: the big bang and big crunch singularity (shell
focusing singularity) at $R(t,r)=0$, and the shell crossing one at
$R'(t,r)=0$. However, if $\frac{M'}{R^{2}R'}$ and $\frac{M}{R^{3}}$
is finite at $R=0$ then there is no shell focusing singularity.
Similarity, if $\frac{M'}{R'}$ is finite at $R'=0$ then there is no
shell crossing singularity. In addition, to get rid of the
unnecessary complications of the shell focusing singularity,
corresponding to a non-simultaneous big bang singularity, we will
assume $t_{n}(r)=0$. This will enable us to concentrate on the
behavior of the collapse of an overdense region in an expanding
universe without interfering with the
complexity of the inherent bang singularity of the metric \cite{man}.\\

\begin{prop} \label{trapped1}There exists a marginally trapped tube for dust cosmological
BH.
\end{prop}
As we know any dynamical BH has collapsing region ($\dot{R}<0$). If
we calculate the expansion for ingoing and outgoing null geodesic
then we get,
$\theta_{(\ell)}\propto(1-\frac{\sqrt{\frac{2M}{R}+f}}{\sqrt{1+f}})$,
$\theta_{(n)}\propto(-1-\frac{\sqrt{\frac{2M}{R}+f}}{\sqrt{1+f}})<0$.
Hence, we see in collapsing region, the expansion for null outgoing
geodesic changes its sign from negative to positive at $R = 2M$ as
we go to larger $R$ and ingoing null geodesics expansion is negative
everywhere. Therefore, the 3-manifold $R = 2M$ is a \emph{marginally
trapped tube} (MTT).

Now we prove that, there is MTT between singularity line
($R=0$,$\eta=2\pi$) and $\dot{R}=0,\eta=\pi$ (boundary between
collapsing and expanding region). It can be seen from (\ref{ltbe1})
that we have $R>2M$ on $\dot{R}=0$ (it is sufficient to put $f>-1$
and $\eta=\pi$). We know that at the singularity (R=0) the
Misner-Sharp mass in non zero $M(r=r_s)=M_{s}>R_s=0$, so if we look
at the $R$ and $2M$ values at $R$ plane, there is a $R_0$ between
$R=0$ and $\dot{R}=0$ that $R_0=2M$ so we have the apparent horizon.
This MTT
can be spacelike, timelike or null surface.\\

\subsection{LTB solutions as a cosmological BH}

 For LTB solutions to be asymptotically FRW certain conditions have
 to be fulfilled. We first note that FRW spaces are special cases of LTB
 metrics: if $R(r,t)$ is separated as $R(r,t) = r a(t)$ we
 obviously get the homogeneous FRW solutions. For the vanishing bang time
 this corresponds to $M(r)=cr^{3}$, $f(r)=-r^{2},0,r^{2}$. Therefore, to
 have an asymptotically open FRW LTB solution  must have $M(r)=cr^{3}$, $f(r)=r^{2}$.

\begin{figure}[h]
\begin{center}
\includegraphics[scale = 0.34]{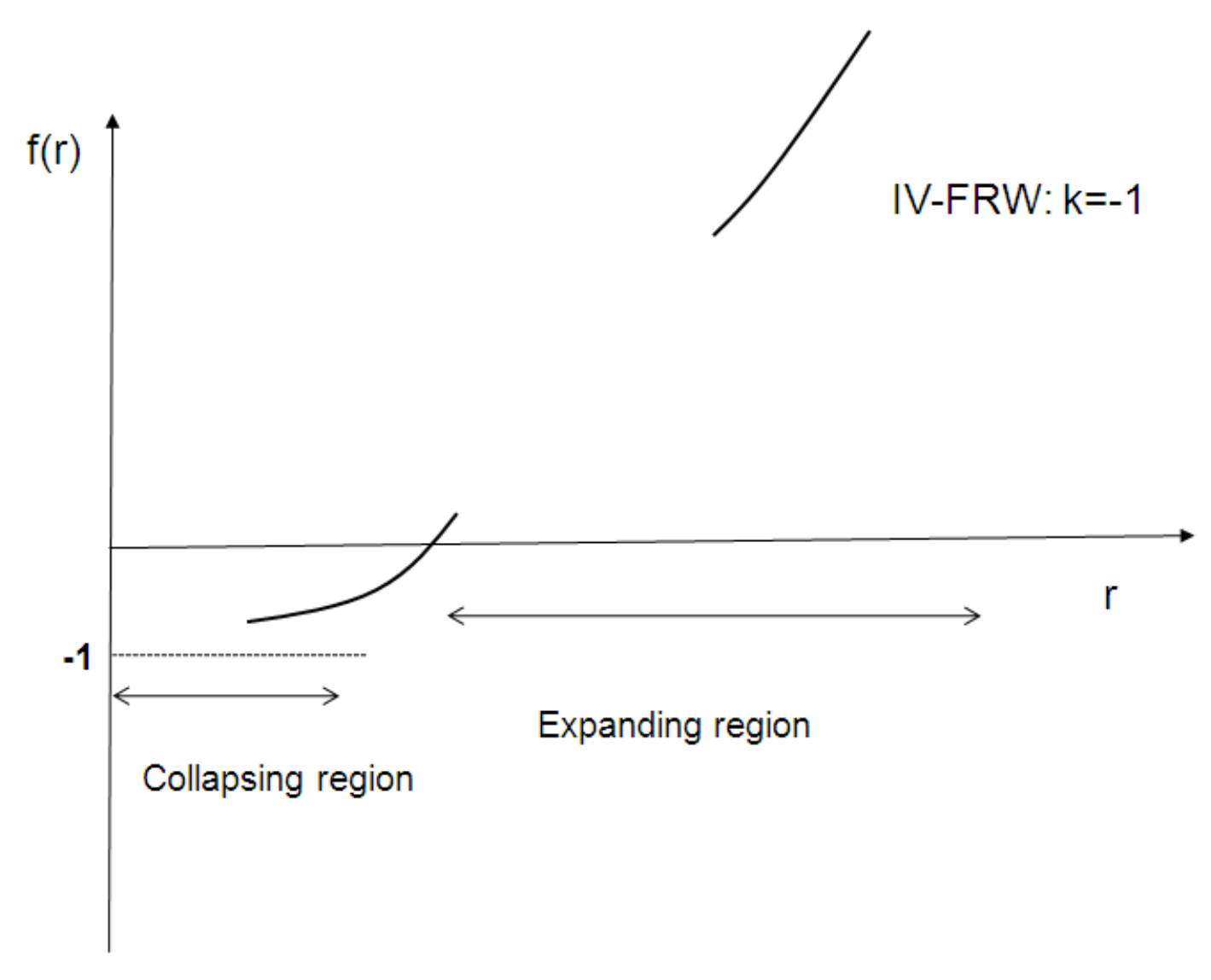}
\hspace*{10mm} \caption{ \label{f(r)}
  Behaviors of the curvature function $f(r)$.}
  \end{center}
\end{figure}

\textbf{Assumptions}: The \emph{first} assumption is that the
Misner-Sharp mass of round spheres increases monotonically concerning comoving coordinate $r$; $M'(r)>0$. Our \emph{second}
the assumption is that we have no shell crossing and $R'>0$ everywhere.
The \emph{third} assumption expresses the idea that wants to look for
conditions leading to an overdense region near the center $r = 0$,
within an expanding universe with $\dot{R}>0$ (at least) far from
the center. However, overdensities in a region around the center
require $\dot{R} < 0$, corresponding to the collapse phase of the
overdense region, which we may assume to start at a time $t_c
> 0$.  We will say that a LTB solution represents a \emph{LTB cosmological BH} if it satisfied these conditions. Finally, equations (\ref{ltbe1}), (\ref{ltbe2}), and
(\ref{ltbe3}) it is easily seen that for the collapsing region one
has to have $f(r)<0$. In contrast, for the universe outside there
must be $f(r)>0$ having expansion for late
time (it can be $f(r)<0$ and we have expanding phase, but after some time this region changes the phase from expanding to collapsing phase).
 Hence there must be at least one root for $f(r)$ (see Fig.\ref{f(r)}).\\

The collapse of the overdense region leads to two new conditions on
the metric coefficients. First, we see from \cite{man}, that at any
constant time shells corresponding to $0<\eta(r)<\pi$ are in an
expanding phase and those corresponding to $\pi<\eta(r)<2\pi$ are in
the collapsing phase.

We will now look for LTB solutions fulfilling these assumptions.

\begin{figure}[h]
\begin{center}
\includegraphics[scale = 0.34]{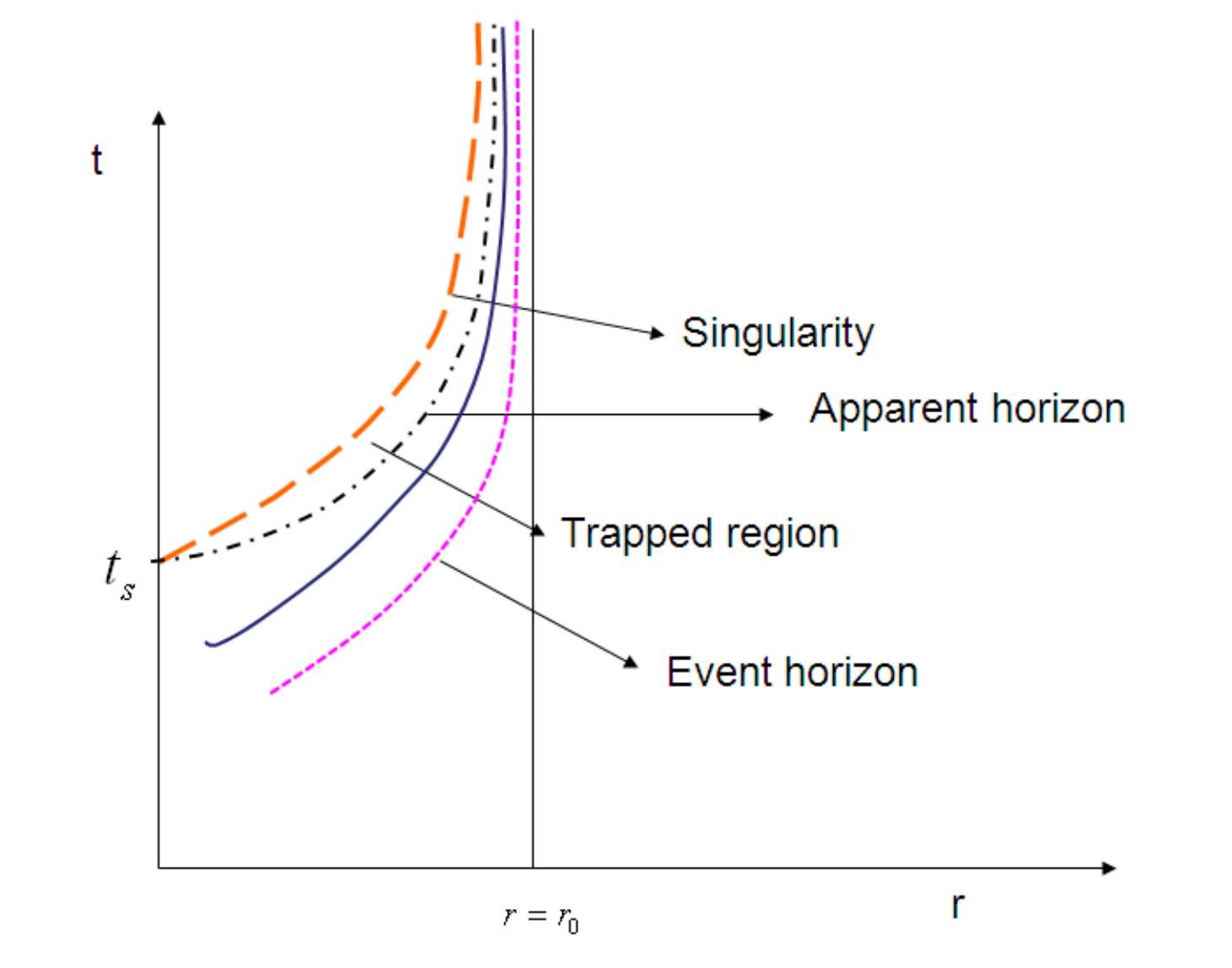}
\hspace*{10mm} \caption{ \label{casual}
  Causal  structure for cosmological BH.}
\end{center}
\end{figure}

\begin{prop}
There is a finite time for LTB cosmological BH at which
$\frac{dr}{dt}|_{MTT}$ becomes positive.
\end{prop}
\textbf{Proof}: For LTB metric we have, \be\label{rprim}
R'=(\frac{M'}{M}-\frac{f'}{f})R
+(-t\dot{R})(\frac{M'}{M}-3\frac{f'}{f}).\ee  In apparent horizon,
$\frac{dr}{dt}|_{AH}=\frac{-\dot{R}}{R'-2M'}$. According to before
proposition. \ref{trapped1} , we have
$\dot{R}|_{AH}=-\sqrt{1+f}\neq0$. We know form the above equation
that\be R'-2M'|_{MTT}=-\frac{f'}{f}R
+(-t\dot{R})(\frac{M'}{M}-3\frac{f'}{f}).\ee The first term on right
hand side can be negative or positive and the second term is +. If
the first term be negative, it is easy to see from above equation
that there will be a finite $t_0$ that$\frac{f'}{f}R<
(-t_0\dot{R})(\frac{M'}{M}-3\frac{f'}{f})$ on MTT (because the
second term is increasing according to $t$  in collapsing region
while the first term remain finite collapsing region).
Hence, $\frac{dr}{dt}|_{AH}>0$.\\

As we see there is a finite time $t_0$ at which $R'-2M'>0$ and so
$R'-M'>0$;
therefor apparent horizon become space like \cite{man}. This is a dynamical horizon which is good candidate for black hole boundary.\\
 We assume  that the collapsing region is limited a finite region
$r_0$ in the expanding region then apparent horizon is limited too.
Consider following proposition from analysis,
\begin{prop}\label{slop}
If $\frac{dr(t)}{dt}>0$ and $r<r_0$, for any $\epsilon_1>0$ and
$\epsilon_2>0$ there will be $t_0$ such that
$\frac{dr}{dt}<\epsilon_1$ or
$\frac{dr}{dt}<\frac{1}{t^{\epsilon_2}}$.

%then at late time($t\mapsto \infty$)
%$\frac{dr}{dt}<\frac{1}{t^{1+\epsilon}}$ which $0<\epsilon<<1$.
\end{prop}
As $t$ becomes larger and larger, the $\frac{dr}{dt}|_{AH}$ tends to
zero. One can see that the $r$ coordinate at event horizon has the
same behavior of singularity line at late time \cite{nielsen}.

\begin{prop}
The singularity line slope in the $t-r$ plane is increasing for cosmological
dust collapse.
\end{prop}
 We know that  $r$ is the comoving radius. We want to construct a
collapsing spherical matter in expanding universe, so we expect that
the singularity forms  at the center, $r=0$ firstly and then any
other shells fall into the singularity. Each shell which is more
nearer to the center will fall into the singularity earlier.
Therefore, the above expression in $t-r$ plane can be mathematically
expresse as $\frac{dt}{dr}|_{singularity}=t'_s(r)>0$. We should note
that $t_s (r)$ and $\dot{R}=0$ have the same behavior in $t-r$ plane
because they correspond to the $\eta=2\pi$ and
$\eta=\pi$ respectively and their difference is only 2 factor.\\
 Now consider the first root of the $f(r=r_0)=0$ which is outside
collapsing region and is in the expanding phase (at any time). It
has finite mass $M$ which corresponds to $\eta=0$. The singularity
line $t_s (r)$ and $\dot{R}=0$ cannot reach to these surface in any
time at  $t-r$ plan. Hence, there is a $r=r_0$ (such that
$f(r=r_0)=0$) that the $t_s (r)$ and
$\dot{R}=0$ curves become asymptote to $r=r_0$ in $t-r$ plan at late time.\\

Although the space time is not asymptotically flat, for the class of
LTB BHs under consideration we can still introduce the notion at an
event horizon as "the very last ray to reach future singularity" or
"the light ray that divides those observers who cannot escape the
future singularity from thus who do" then we have the following
result.

\begin{prop}\label{trapped} Every trapped surface ($T$) for cosmological BH is located inside the event
horizon.
\end{prop}

Our matter is dust and satisfies energy conditions. Suppose that
there is a point $p$ that is located outside the event horizon.
Therefore, it can send a ray to point $q$ infinity (not singularity).
Hence, according to theorem 9.3.11 at \cite{waldbook}  there is null
geodesic $\gamma$ from $p$ to $q$ is orthogonal to $T$ has no
conjugate point between $T$ and $q$. However, this is impossible,
because according to theorem 9.3.6, $\gamma$ must have a conjugate
point within affine parameter $2/|\theta_0|$ from $p$, where
$\theta_0<0$ is the expansion at $p$ of the orthogonal null geodesic
congruence from $T$ to which
$\gamma$ belongs.\\
Geometry of the dynamical horizon $H$ is represented by The unit
normal to $H$ by $\hat{\tau^a}$; $g_{ab}\hat{\tau^a}\hat{\tau^b}=-1$
. The unit space-like vector orthogonal to $S$ and tangent to $H$ is
denoted by $\hat{r^a}$. Finally, we will fix the rescaling freedom
in the choice of null normals via $l^a=\hat{\tau^a}+\hat{r^a}$ and
$n^a=\hat{\tau^a}-\hat{r^a}$. As is usual in general relativity, the
notion of energy is tied to a choice of a vector field. The
definition of a dynamical horizon provides a preferred direction
field; that is along $\ell^a$. To fix the proportionality factor, or
the lapse $N$, let us first introduce the area radius $R$, a
function  which is constant on each $S$ and satisfies $a_S = 4\pi
R^2$. Since we already know that area is monotonically increasing,
$R$ is a good coordinate on $H$. Now, the 3-volume $d^3V$ on $H$ can
be decomposed as $d^3V = |\partial R|^{-1}dR d^2V$ where $\partial$
denotes the gradient on $H$. Therefore, as we will see, our
calculations will simplify if we choose $N_R = |\partial R|$.

Fix two cross sections $S_1$ and $S_2$ of $H$ and denote by $\Delta
H$ the portion of $H$ they bound. We are interested in calculating
the flux of energy associated with $\xi_{(R)}^a = N_R \ell^a$ across
$\Delta H$. Denote the flux of \emph{matter} energy across $\Delta
H$ by $\mathcal{F}^{(R)}_{\rm matter}$:
\begin{eqnarray} \mathcal{F}^{(R)}_{\rm matter} := \int_{\Delta H}
T_{ab}\hat{\tau^a}\xi_{(R)}^b d^3V=\nonumber\\
\frac{1}{G}(M(r_2)-M(r_1)).\end{eqnarray}

 LTB BHs, dynamical horizon behaves like
(\ref{casual}; at  late time, matter flux become zero and the
horizon radius become fixed at radius $R=2M(r_0)$ where $M(r_0)$ is
Misner-Sharp mass in the horizon. For more practical case we can
calculate the matter accretion according time along the dynamical
horizon by \be
\frac{dM(r)}{dt}|_{DH}=\frac{dM(r)}{dr}\frac{dr}{dt}|_{DH}.\ee To
show that the dynamical horizon becomes isolated horizon, we have to
show that the dynamical horizon become Slowly evolving horizon at
late time. It can be seen from \cite{man} that
\begin{equation}
\frac{dt}{dr}|_{AH}=\frac{R' -2 M'}{-\dot{R}}=\frac{R' -2M'}{1+f}.
\end{equation}
We know $R'>0$ has regular behavior (no shell crossing) and $f(r)$
is finite along the MTT. From Proposition.\ref{slop} we $R'
-2M'|_{AH}$ becomes large and $R' -M'|_{AH}$ do too. We follow the
Booth \cite{booth06} calculations to show that our dynamical horizon
becomes a \emph{slowly evolving horizon} at a late time. In defining a
slowly evolving horizon, it is convenient to further restrict the
scaling of the null vectors. To do this we label the foliating MTSs
with a parameter $\upsilon$ and choose the scaling and an evolution
parameter $C$ so that \be V^\mu=\ell^\mu-c n^\mu \ee is tangent to
DH and \be \pounds_V \upsilon=1.\ee
According to \cite{booth06}, if we calculate $c$ and $\epsilon$
along the DH for LTB metric, we have \be
c=2\sqrt{1+f}\frac{M'}{R'-M'}|_{AH}\ee,  \be \epsilon^2=8\sqrt{1+f}
\frac{M'}{R'-M'}|_{AH}\ee if $c,\epsilon<<1$, then slowly evolving
horizons conditions become satisfied. As we see in above , $R'
-M'|_{AH}$ becomes large at large $t$, therefore our DH becomes
slowly evolving horizons at a late time. We can see from the Einstein equation that the density become small around the apparent horizon
($R$ and $M'$ are finite because we have no singularity at there)
and we will have a void around the apparent horizon at a late time.

\subsection{Redshift}

If an emitter sends a light ray to an observer with a null vector
$k^{\mu}$, the relative light redshift that is calculated by
an observer with 4-velocity $u^{\mu}$ is,

\begin{equation}
1+z=\frac{(k_\mu u^\mu)_e}{(k_\mu u^\mu)_o}.
\end{equation}

From \cite{dwivedi}, we can can calculate the redshift for observer
who sit at $r=const$ with $u^\mu=(1,0,0,0)$ as below,
\begin{eqnarray}
k^t=c_0 ~ exp \left(-\int \frac{\dot{R}'}{\sqrt{1+f}}dr \right) =\nonumber\\
c_0~ exp(\int
\frac{-1}{\sqrt{1+f}}\left(\frac{M'}{R\dot{R}}-\frac{MR'}{\dot{R}R^{2}}+\frac{f'}{2\dot{R}})
dr \right).
\end{eqnarray}
As we saw in the last section, $R'$ becomes large for a late time in
the apparent horizon and $(k_\mu u^\mu)_e$ will have a larger value as $t$
become larger. Therefore, the light that comes out from this surface
to the external observer will be fainted and has an infinite redshift
relative to the distant observer at large time.\\

\section{Cosmological BH with perfect fluid}

In the last section, we limited ourselves to a dust source and studied it at
BH formation in expanding background. In this section, we want to
probe if the same properties of the dust cosmological BH collapse
are held for general perfect fluid collapse. We assume perfect fluid
with an equation of state of the form $p=w\rho$. If $w=0$ the fluid
becomes dust. We now assume that $w\neq0$ and has a finite value. Take
a collapsing ideal fluid within a compact spherically symmetric
spacetime region described by the following metric in the comoving
coordinates $(t,r,\theta,\varphi)$:
\begin{equation}
ds^{2}=-e^{2\nu(t,r)}dt^{2}+e^{2\psi(t,r)}dr^{2}+R(t,r)^{2}d\Omega^{2}.
\end{equation}
Assuming the energy-momentum tensor for the perfect fluid in the
form
\begin{eqnarray}
 T^{t}_{t}=-\rho(t,r),~~T^{r}_{r}=p_{r}(t,r),~~T^{\theta}_{\theta}=\nonumber\\
 T^{\varphi}_{\varphi}=p_{\theta}(t,r)=w \rho(t,r),
\end{eqnarray}
with the week energy condition
\begin{equation}
 \rho\geq0~~\rho+p_{r}\geq0~~\rho+p_{\theta}\geq0,
\end{equation}
where $w$ is constant. Einstein equations give,

 \ba \label{gltbe2}
 \rho=\frac{2M'}{R^{2}R'}~,~~p_{r}=-\frac{2\dot{M}}{R^{2}\dot{R}},
\ea
\begin{equation}
 \nu'=\frac{2(p_{\theta}-p_{r})}{\rho+p_{r}}\frac{R'}{R}-\frac{p'_{r}}{\rho+p_{r}},
\end{equation}
\begin{equation}
 -2\dot{R}'+R'\frac{\dot{G}}{G}+\dot{R}\frac{H'}{H}=0,
\end{equation}
where
\begin{equation}
G=e^{-2\psi}(R')^{2}~~,~~H=e^{-2\nu}(\dot{R})^{2},
\end{equation}
and $M$ is defined by

\ba \label{gltbe3} G-H=1-\frac{2M}{R}. \ea

The function $M$ can also be written as
\begin{eqnarray} \label{ms-def}
M=\frac{1}{2}\int_{0}^{R}\rho R^{2}dR,
\end{eqnarray}
or
\begin{equation}
M=\frac{1}{8\pi}\int_{0}^{r}\rho
\sqrt{(1+(\frac{dR}{d\tau})^{2}-\frac{2M}{R})}d^{3}V,
\end{equation}
where
\begin{equation}
d^{3}V=4\pi e^{\psi}R'dr,
\end{equation}
and
\begin{equation}
\frac{d}{d\tau}=e^{-\nu}\frac{d}{dt}.
\end{equation}
The last form of the function $M$ indicates that when considered as
energy, it includes a contribution from the kinetic energy and the
gravitational potential energy. $M$ is called the Misner-Sharp
energy.\\
 Hayward \cite{hay-ms} showed that in the Newtonian limit of
a perfect fluid, $M$ yields the Newtonian mass to leading order and
the Newtonian kinetic and potential energy to the next order. In
a vacuum, $M$ reduces to the Schwarzschild energy. At null and spatial
infinity, $M$ reduces to the Bondi-Sachs and Arnowitt-Deser-Misner
energies respectively \cite{hay-ms}. Similar to dust case the flux
of \emph{matter} energy across $\Delta H$ by $\mathcal{F}^{(R)}_{\rm
matter}$:
\begin{eqnarray} \mathcal{F}^{(R)}_{\rm matter} := \int_{\Delta H}
T_{ab}\hat{\tau^a}\xi_{(R)}^b d^3V=\nonumber\\
\frac{1}{G}(M(t_2,r_2)-M(t_1,r_1)).\end{eqnarray}
\\

We assume $M'(t,r)>0$ and $\dot{M}(t,r)>0$ in collapsing region
which came form positive sign of density and pressure.\\
\emph{Existence of marginally trapped tube for cosmological BH:} As
we know any dynamical BH has collapsing region ($\dot{R}<0$). If we
calculate the expansion for ingoing and outgoing null geodesic then
we get, $\theta_{(\ell)}\propto(1-\sqrt{1
+\frac{\frac{2M}{R}-1}{G}})$, $\theta_{(n)}\propto(-1-\sqrt{1
+\frac{\frac{2M}{R}-1}{G}})<0$. Hence, we see in collapsing region,
the expansion for null outgoing geodesic changes its sign from
negative to positive at $R = 2M$ as we go to larger $R$ and ingoing
null geodesics expansion is negative everywhere. Therefore,
3-manifold $R = 2M$ is \emph{marginally trapped tube}.

For proof that there exist MTT between the singularity line and the
the boundary between expansion region and collapsing region, we know
that at $\dot{R}=0$ so $H=0$ and from (\ref{gltbe3}) we have $G>0$
then $R>2m|_{\dot{R}=0}$. On the other hand in singularity
$2M(t,r_s)>R=0$. If we look at the $R$ and $2M$ values at the $R$ plane,
there is a $R_0$ between $R=0$ and $\dot{R}=0$ that $R_0=2M$ so we
have the
MTT.\\

\begin{figure}[h]
\begin{center}
\includegraphics[scale = 0.34]{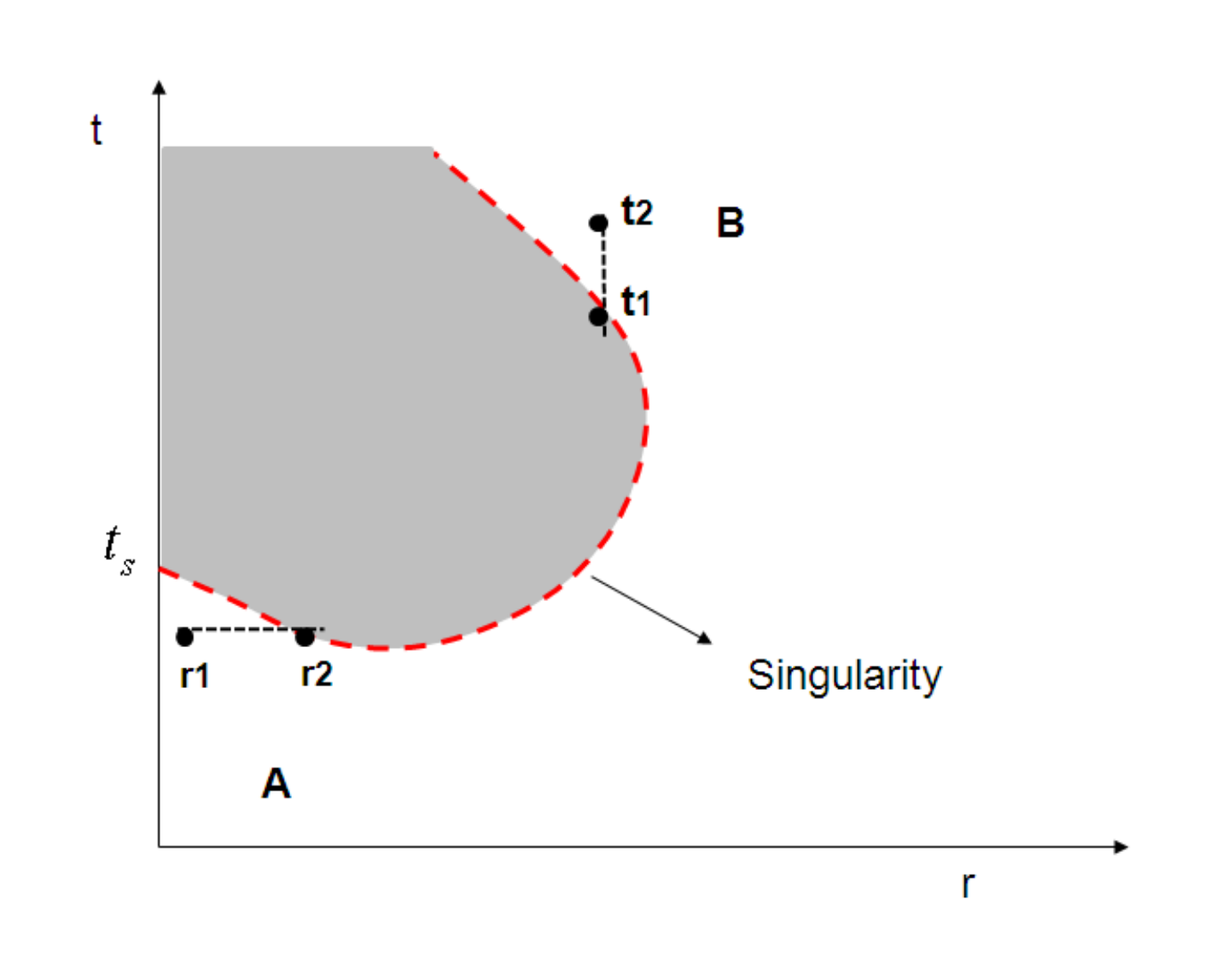}
\hspace*{10mm} \caption{ \label{inexample}
  Impossible shape for the singularity line in BH collapse.}
\end{center}
\end{figure}

Similarly, we assume that $M'>0$ (which means that Misner-Sharp mass
is increasing ), we have no shell crossing $e^{2\psi(t,r)}>0$(or
equivalently for finite $G$ $R'>0$), $\dot{M}>0$ in collapsing
region and $\dot{M}<0$ in expanding region. In the region between
collapsing and expanding matter; $\dot{R}=0$ and then $\dot{M}=0$
because from (\ref{gltbe2}) we have no singularity in pressure.
These condition are compatible with $\rho$ and $p$ positive sign.\\
\begin{prop}
the Singularity line slope in $t-r$ plane increases for general
spherical fluid collapse.
\end{prop}
We want to show if we assume that singularity line has another shape
for example part $A$ and part $B$ ( see Fig.\ref{inexample}), then
we get contradictions with our assumptions. Consider part $A$, if
these case exist then we have $R(r_1)>R(r_2)=0$ which is in
contradiction with $R'>0$ and Consider part $B$ if these cases exist
then we have $R(t_1)>R(t_2)=0$ which is in contradiction with
$\dot{R}<0$. Therefore,  $t'_s>0$ comes from the above assumptions
for cosmological BH collapse.

Cosmological expansion assumption forces singularity to be limited in
the part of the space time around the center, and the other part
expands with cosmological expansion. Hence, there is an event horizon
that separates causal geodesics that fall into the singularity from
other geodesics. This event horizon is asymptotic to the singularity
at a late time. Similar to the dust case, the trapped region is inside
the event
horizon see Proposition.\ref{trapped}.\\
\begin{prop}\label{} The apparent horizon slope in $t-r$ plane becomes infinite at late time.
\end{prop}
We know that the singularity slop is strictly positive and
singularity is limited in a finite $r$, therefor its slop becomes
infinite (see Proposition.\ref{slop}). The apparent horizon is
among the singularity line and the event horizon, so as even horizon
tends to the singularity then apparent horizon tends to singularity
and it becomes asymptotic to the singularity
at late time and its slop becomes infinite.\\
We want to show that MTT becomes slowly evolving horizons large
time. We know the expression
\begin{equation}
\frac{dt}{dr}|_{AH}=\frac{R' -2 M'}{2\dot{M}-\dot{R}},
\end{equation}
is large at from above proposition late time. We know $R'>0$ has
regular behavior (no shell crossing for finite $G$), $\dot{R}<0$ and
$\dot{M}\geq 0$ are finite along the apparent horizon. Therefor, we
can infer $2\dot{M}-\dot{R}$ is finite. Since the density is finite
around the horizon,
 so $R' -2 M'=R' -2 R'\rho R^2$ should be large and then  $R'|_{AH}\gg0$ becomes large.\\
If we want than MTT becomes space like, we should have
\begin{equation}
 \frac{\frac{dt}{dr}|_{AH}}{\frac{dt}{dr}|_{null}}=-1< \left( \frac{1 - \frac{2 M'}{R'}}{1 - \frac{2 \dot{M}}{\dot{R}}} \right)<1.
\end{equation}
Because of $\dot{M}|_{AH}>0$ and $\dot{R}|_{AH}<0$, the above
condition will be satisfied and MTT becomes dynamical
horizon at large time.\\

Similar to the LTB case, it can be shown that the dynamical horizon
satisfying \emph{slowly evolving horizons} at large time. According
to \cite{booth06}, if we calculate $c$ and $\epsilon$ along the DH
for perfect fluid metric, we have \be c=2\frac{M'\dot{R}+w
M'\dot{R}}{M'\dot{R}-w M'\dot{R}-\dot{R}R'}|_{AH}\ee, \be
\epsilon^2=8 \frac{M'\dot{R}+w M'\dot{R}}{M'\dot{R}-w
M'\dot{R}-\dot{R}R'}|_{AH},\ee which we put $\dot{M}=-w
M'\frac{\dot{R}}{R'}$ from the perfect fluid assumption. If
$c,\epsilon<<1$, then slowly evolving horizons conditions become
satisfied. As we calculate above, $R'|_{AH}$ become large at
large $t$, therefor the dynamical horizon becomes slowly evolving
horizons at large time. Similar to LTB, the density becomes small
around the  ($R$ and $M'$ are finite around the MTT) and void will
be formed around the dynamical horizon at a late time.\\

\section{Cosmological BH with cosmological constant}
One of the characteristic properties of a black hole is its horizon.
One can show that cosmological constant can affect the
nature of the singularity and final fate of the black hole
\cite{joshi-lamda}. But we want to show that the local behavior of the
gravitational collapse is not affected by the cosmological constant as
a dark energy candidate. If we calculate the expansion for ingoing
and outgoing null geodesic then we get,
$\theta_{(\ell)}\propto(1-\frac{\sqrt{\frac{\Lambda
R^2}{3}+\frac{2M}{R}+f}}{\sqrt{1+f}})$,
$\theta_{(n)}\propto(-1-\frac{\sqrt{\frac{\Lambda
R^2}{3}+\frac{2M}{R}+f}}{\sqrt{1+f}})<0$. So the apparent horizon
surface change from $2M=R$ to $2M=R-\frac{\Lambda R^3}{3}$ surface
with
the same properties. \\
For general perfect fluid we have the same change in apparent
horizon surface from expansion of light ray,
$\theta_{(\ell)}\propto(1-\sqrt{1 +\frac{\frac{\Lambda
R^2}{3}+\frac{2M}{R}-1}{G}})$, $\theta_{(n)}\propto(-1-\sqrt{1
+\frac{\frac{\Lambda R^2}{3}+\frac{2M}{R}-1}{G}})<0$ (The apparent
horizon surface $2M=R-\frac{\Lambda R^3}{3}$).

\be \mathcal{F}^{(R)}_{\rm matter} := \int_{\Delta H}
(T_{ab}-\frac{\Lambda}{8\pi G}g_{ab})\hat{\tau^a}\xi_{(R)}^b
d^3V.\ee

It can be seen from \cite{ashtekar02} that the area law becomes, \be
dE^t=\frac{\tilde{\kappa}_R}{8 \pi G}(1-\Lambda R^2)d(4 \pi R^2) \ee
which $\kappa_R=\frac{1}{2R}$ is effective surface gravity. We can
integrate from the above equation  \be E^t(R)|_{AH}=\frac{1}{2
G}(R-\frac{\Lambda R^3}{3})=\frac{M(r)}{G}. \ee If we compare a
cosmological BH with a radius $R_0$ in de Sitter background
($\Lambda>0$) with non de Sitter background, we will get that de
Sitter background black hole will have less Misner-Sharp mass $M$ in
black hole with radius $R_0$.\\
The local property of the black hole evolution will not change with
adding the cosmological constant term because if we put cosmological
constant realistic value in the Einstein equation with $\Lambda
\ll10^{-35} s^{-2}$ and mass $M=10^{12}M_\odot$, we get
\begin{equation}
G-H=1-\frac{2M}{R}+\frac{\Lambda}{3}R^2,~~~~\frac{\Lambda}{3}R^2\ll\frac{2M}{R},
\end{equation}
$\frac{\Lambda}{3}R^2<10^{-20}$ . Hence, cosmological constant don't
affect on local black hole physics.\\

\section{DISCUSSION}

Considering the back hole as a dynamical object in expanding
background shows new properties that a Schwarzschild back hole is
not able to show. In particular, the mass and the flux of matter can
have different properties in expanding background. Recent analysis
has shown that these dynamical BHs have the flexibility to explore some
basic questions such as cosmic censorship conjecture, quasi-local
definition masses and BH thermodynamic in non-asymptotic flat background \cite{ashtekar02, cencorship}. Thus, the first attempt is
making a good model for these BH (cosmological BH) with a reasonable
kind of matter \cite{man}. In this paper, we assumed some acceptable
properties for these BH and consider its singularity and its
horizons behavior along the time. We found that the singularity line
 shows a special behavior and MTT becomes dynamical horizon as a
candidate for BH boundary in non-asymptotic flat background.
Furthermore, in terms of global behavior in time, Misner-Sharp mass
has this ability to describe the mass of the cosmological BH and
its time evolution across the dynamical horizon becomes the fluid
accretion. In contrast to the Schwarzschild BH, the light with infinite
redshift cannot come from the apparent horizon to distance observer.
We showed that for the dust cosmological collapse, the apparent
horizon becomes infinite redshift surface only at the late time for
distance observer. Finally, we found that we can neglect the
cosmological constant (as a dark energy candidate) for the local
behavior of the BH in time.\\

\section{ACKNOWLEDGMENT}
I am particularly indebted to Abhay Ashtekar for his critical
reading and many discussions throughout this paper. I also would like
to thank Erfan Salavati and Reza Mansouri for their helpful discussions.
This research was supported in part by NSF grant PHY0854743, the
George A. and Margaret M. Downsbrough Endowment, the Eberly research
funds of Penn State.


\begin{thebibliography}{99}
\bibitem{sultana}
J. Sultana and C.C. Dyer, Gen. Rel. Grav. 37, 1349 (2005).
\bibitem{McVittie}
G.C. McVittie, Mon. Not. R. Astr. Soc. 93, 325 (1933).

\bibitem{Einstein Straus}
A. Einstein and E.G. Straus, Rev. Mod. Phys. 17, 120 (1945); 18, 148


\bibitem{LTB}
R. C. Tolman, Proc. Natl. Acad. Sci. U.S.A. 20, 410 (1934); G.
Lemaitre, Ann. Soc. Sci. Bruxelles I A53, 51 (1933); H. Bondi,
Mon. Not. R. Astron. Soc. 107, 343 (1947).

\bibitem{cencorship}
P. S. Joshi and T. P. Singh, Phys. Rev. D 51, 6778 (1995); R. P. A.
C. Newman, Class. Quantum Grav. 3, 527 (1986); D. Christodoulou,
Commun. Math. Phys. 93, 171 (1984); D. M. Eardly and L. Smarr, Phys.
Rev. D 19, 2239 (1979).
\bibitem{joshi} P. S. Joshi and I. H. Dwivedi, Phys. Rev. D 47, 5357
(1993); S Jhingan, PS Joshi(gr-qc/ 9701016); A. Chamorro, S.S.
Deshingkar, I.H. Dwivedi, and P.S. Joshi, Phys. Rev. D 63, 084018
(2001).

\bibitem{man}
J.~T.~Firouzjaee and R.~Mansouri,
%``Asymptotically FRW black holes,''
Gen. Rel. Grav. \textbf{42}, 2431-2452 (2010).

\bibitem{manmass}
J.~T.~Firouzjaee, M.~P.~Mood and R.~Mansouri,
%``Do we know the mass of a black hole? Mass of some cosmological black hole models,''
Gen. Rel. Grav. \textbf{44}, 639-656 (2012),
[arXiv:1010.3971 [gr-qc]].


\bibitem{waldbook}
Robert M. Wald, \emph{General Relativity} (University of Chicago
Press, Chicago, 1984).

\bibitem{Hayward94}
S. A. Hayward, Phys. Rev. D 49, 6467 (1994).
\bibitem{hay-ms}
S. A. Hayward, Phys. Rev. D 53, 1938 (1996).
\bibitem{ashtekar99}
A. Ashtekar, C. Beetle, O. Dreyer, S. Fairhurst, B. Krishnan, J.
Lewandowski and J. Wisniewski, Phys. Rev. Lett. 85, 3564-3567
(2000).

\bibitem{ashtekar02}
A. Ashtekar and B. Krishnan, Phys. Rev. Lett. 89, 261101 (2002); A.
Ashtekar and B. Krishnan, Phys. Rev. D 68, 104030 (2003).

\bibitem{booth04}
Booth I. and Fairhurst S., Phys. Rev. Lett. 92, 011102 (2004).

\bibitem{kra-hel-sf}
Interactions K. Bolejko, A. Krasinski, C. Hellaby and M-N,
\emph{Structures in the Universe by Exact Methods - Formation,
Evolution} (Cambridge University Press, 2009); C. Hellaby and A.
Krasinki , Phys. Rev. D 73, 023518 (2004).


\bibitem{kra-hel-BH}
A. Krasinki and C. Hellaby, Phys. Rev. D 69, 023502 (2004).



\bibitem{harada}
Tomohiro Harada, C. Goymer and B.J. Carr,  Phys. Rev. D66, 104023;
H.~Saida, T.~Harada and H.~Maeda,
  %``Black Hole Evaporation in an Expanding Universe,''
  Class.\ Quant.\ Grav.\  {\bf 24}  4711 (2007).


\bibitem{nielsen}
Alex B. Nielsen [gr-qc/1006.2448].



\bibitem{booth06}
William Kavanagh, Ivan Booth, Phys. Rev. D 74 044027 (2006).

\bibitem{dwivedi}
I. H. Dwivedi Phys. Rev. D 58, 064004 (1998).


\bibitem{joshi-lamda}
T. Arun Madhav, Rituparno Goswami, and Pankaj S. Joshi, Phys. Rev. D
72, 084029 (2005); E.Babichev, V.Dokuchaev and Y.Eroshenko, Phys. Rev.
Lett. 93, 021102 (2004)



\end{thebibliography}
\end{document}